\documentclass{webofc}
\usepackage[varg]{txfonts}   
\usepackage{graphicx,bm}
\begin{document}
\title{Signatures for tetraquark mixing from partial decay widths of the two light-meson nonets~\footnote{This talk is mainly based on Ref.~\cite{Kim:2022qfj}.}}
%
%

\author{\firstname{Hungchong} \lastname{Kim}\inst{1}\fnsep\thanks{\email{bkhc5264@korea.ac.kr}} \and
        \firstname{K. S.} \lastname{Kim}\inst{2}\fnsep
}

\institute{Center for Extreme Nuclear Matters, Korea University, Seoul 02841, Korea
\and
           School of Liberal Arts and Science, Korea Aerospace University, Goyang, 412-791, Korea
          }

\abstract{%
In this talk, we present successful aspects of the tetraquark mixing model
for the two light-meson nonets in the $J^{PC} = 0^{++}$ channel,
the light nonet [$a_0(980)$, $K_0^*(700)$, $f_0(500)$, $f_0(980)$] and the heavy nonet
[$a_0(1450)$, $K_0^*(1430)$, $f_0(1370)$, $f_0(1500)$].
In particular, we focus on how their experimental partial decay widths extracted from Particle Data Group (PDG)
can support this mixing model.
Currently, the experimental data exhibit an unnatural tendency
that partial widths of the light nonet are consistently larger than
those of the heavy nonet. This unnatural tendency can be explained if the coupling into
two pseudoscalar mesons is enhanced in the light nonet and suppressed in the heavy nonet as predicted by the tetraquark mixing model.
Therefore, this could be strong evidence to support for the tetraquark mixing model.
}
\maketitle
%
In PDG~\cite{PDG22}, there are two nonets in the $J^{PC}=0^{++}$ channel: the light nonet [$a_0(980)$, $K_0^*(700)$, $f_0(500)$, $f_0(980)$]
and the heavy nonet [$a_0(1450)$, $K_0^*(1430)$, $f_0(1370)$, $f_0(1500)$].
About six years ago, we proposed a tetraquark mixing model
that treats the two nonets as tetraquarks generated by mixing the two tetraquark types, $|000\rangle$ and $|011\rangle$~\cite{Kim:2016dfq,Kim:2017yur,Kim:2017yvd,Kim:2018zob,Lee:2019bwi}.
The first type, $|000\rangle$, represents the spin-0 tetraquarks formed by combining the spin-0 diquark
of the structure ($\bar{\bm{3}}_c, \bar{\bm{3}}_f$) and its antidiquark.
The second type, $|011\rangle$, represents the spin-0 tetraquarks constructed by the spin-1 diquark of the structure ($\bm{6}_c, \bar{\bm{3}}_f$) and its antidiquark.
Their mixtures that diagonalize
the color-spin interaction, $V_{CS}$, have been identified as two physical nonets,
\begin{eqnarray}
|\text{Heavy~nonet} \rangle &=& -\alpha | 000 \rangle + \beta |011 \rangle \label{heavy}\ ,\\
|\text{Light~nonet} \rangle~ &=&~~\beta | 000 \rangle + \alpha |011 \rangle \label{light}\ .
\end{eqnarray}
The mixing parameters are $\alpha\approx \sqrt{{2}/{3}}$, $\beta\approx 1/\sqrt{3}$ also fixed by the diagonalization process.

This model has been tested in several occasions~\cite{Kim:2016dfq,Kim:2017yur,Kim:2017yvd,Kim:2018zob,Lee:2019bwi}
and appears to have successful aspects, such as qualitatively explaining the mass of the two nonets
and the difference in mass between the two nonets.
Specifically, the mixing creates a large hyperfine mass for the light nonet, on the order of $-500$ MeV,
which can explain why the light nonet has masses less than 1 GeV even though its members are composed of 4 constituent quarks.
The mixing makes a hyperfine mass for the heavy nonet to around $\sim -20$ MeV, which can
explain qualitatively why the heavy nonet has masses not far from $4m_q$, four times of the constituent quark mass.
More interestingly,  the mass difference between the two nonets, which is around 500 MeV or more, can be explained by the
the hyperfine mass difference, $\Delta M \approx \Delta \langle V_{CS}\rangle$.
It all comes from the fact that the two types of tetraquark mix together to create the two nonets.
\begin{table}
\caption{Decay modes and the coupling strengths of the light nonet (heavy nonet), $G$ ($G^\prime$), calculated from the tetraquark mixing model.
}
\label{coup}
\centering
\begin{tabular}{l|c|l|c}  \hline\hline
\multicolumn{2}{c|}{Light nonet} & \multicolumn{2}{c}{Heavy nonet} \\
\hline
     Decay mode &  $G$  & Decay mode & $G^\prime$ \\
\hline
$~a_0(980)\rightarrow \pi \eta$  &$0.6076$ & $~a_0(1450)\rightarrow \pi \eta$&  $ 0.1406$  \\
$~a_0(980)\rightarrow K\bar{K}$  &0.7441 & $~a_0(1450)\rightarrow K\bar{K}$ &  $0.1722$ \\
\hline
$~K_0^*(700)\rightarrow \pi K$ & $0.5253$  & $~K_0^*(1430)\rightarrow \pi K$ &  $0.1251$   \\
\hline
$~f_0(500)\rightarrow \pi \pi$ & $-0.3310$  & $~f_0(1370)\rightarrow \pi \pi$ &  $-0.0785$   \\
\hline
$~f_0(980)\rightarrow \pi \pi$ & $-0.1690$  & $~f_0(1500)\rightarrow \pi \pi$ &  $-0.0394$   \\
$~f_0(980)\rightarrow K\bar{K}$ & $-0.4685$  & $~f_0(1500)\rightarrow K\bar{K}$ & $-0.1093$  \\
\hline
\end{tabular}
\end{table}

The most striking prediction of the tetraquark mixing model is that the coupling strengths of the two nonets decaying into two pseudoscalar
mesons must satisfy the following inequality~\cite{Kim:2017yur,Kim:2022qfj}
\begin{eqnarray}
|G|~\text{(light~nonet)} \gg |G^\prime|~\text{(heavy~nonet)} \label{gg}\ .
\end{eqnarray}
It says that the coupling strength ($G$) of the light nonet is much larger than that ($G^\prime$) of the heavy nonet.
To illustrate this prediction, we notice that tetraquarks ($q_1 q_2 \bar{q}_3\bar{q}_4$) either in $|000\rangle$ or in $|011\rangle$
are composed of diquark ($q_1q_2$) and antidiquark ($\bar{q}_3\bar{q}_4$) with definite color, spin, and flavor states.
If the tetraquark is recombined into quark-antiquark pairs, $q_1\bar{q}_3$, $q_2\bar{q}_4$,
its wave function in color space has a component consisting of two color-singlets and additional component of two color-octets.
This can be schematically written as,
\begin{eqnarray}
[q_1 q_2 \bar{q}_3\bar{q}_4]_{\bm{1}_c} \sim (q_1 \bar{q}_3)_{\bm{1}_c}\otimes (q_2 \bar{q}_4)_{\bm{1}_c} +
[(q_1 \bar{q}_3)_{\bm{8}_c}\otimes (q_2 \bar{q}_4)_{\bm{8}_c}]_{\bm{1}_c} \label{recomb}\ .
\end{eqnarray}
Tetraquarks can decay into two pseudoscalar mesons through the first component containing two color-singlets. Of course, to make two-meson modes,
the tetraquarks has to be recombined in spin and flavor space also.
The coupling strengths of our concern, which are basically the coefficients of the two-meson modes,
can be calculated from the recombination factors from color, spin and flavor.
The tetraquark type, $|000\rangle$, can decay into two mesons through this component and the other type, $|011\rangle$,
can decay into the same mesons through this component also.
However, due to the opposite signs in the heavy nonet, Eq.~(\ref{heavy}), two-meson modes from $|000\rangle$ and
$|011\rangle$ partially cancel out to suppress the coupling strengths, $|G^\prime|~\text{(heavy~nonet)}$.
On the other hand, the two-meson modes have the same sign in the light nonet, Eq.~(\ref{light}), and they add up each other to
enhance the coupling strengths, $|G|~\text{(light~nonet)}$.  This leads to the inequality like Eq.~(\ref{gg}).
A more detailed explanation can be found also in Refs.~\cite{Kim:2017yur,Kim:2022qfj}.

Table~\ref{coup} shows the coupling strengths obtained by recombining the wave functions
of Eqs.~(\ref{heavy}),(\ref{light}) in terms of quark-antiquark pairs.
As advertised in Eq.~(\ref{gg}), we clearly see that the tetraquark mixing model predicts that $|G|$ is much larger than $|G^\prime|$.
More interestingly, this prediction can be experimentally verified by examining the partial decay width
because the partial decay width is given as
\begin{eqnarray}
\Gamma_{partial} = (\text{coupling strength})^2 \Gamma_{kin} \label{partial}\ .
\end{eqnarray}
Here, the kinematical decay width, $\Gamma_{kin}$, which depends only on kinematical factors in decay processes, should satisfy
\begin{eqnarray}
\Gamma_{kin} \text{(light~nonet)} \ll \Gamma_{kin} \text{(heavy~nonet)} \label{kin}\ ,
\end{eqnarray}
because the heavy nonet has much more phase space for its decay than the light nonet.
By equating Eq.~(\ref{partial}) with the experimental partial decay widths, one can extract some information
of the coupling strengths, $G, G^\prime$, of the two nonets decaying into two pseudoscalar mesons.

\begin{table}
\centering
\caption{Partial decay modes and their widths that are extracted from PDG 2022~\cite{PDG22}.
}
\label{partial width}
\begin{tabular}{l|c|l|c}  \hline\hline
\multicolumn{2}{c|}{Light nonet} & \multicolumn{2}{c}{Heavy nonet} \\
\hline
     Decay mode &  $\Gamma_{exp}$(MeV)  & Decay mode & $\Gamma_{exp}$(MeV) \\
\hline
$~a_0(980)\rightarrow \pi \eta$  &$60$ & $~a_0(1450)\rightarrow \pi \eta$&  $15.4\text{--}20.5$  \\
$~a_0(980)\rightarrow K\bar{K}$  &10.6 & $~a_0(1450)\rightarrow K\bar{K}$ &  $13.5\text{--}18.0$ \\
\hline
$~K_0^*(700)\rightarrow \pi K$ & $468$  & $~K_0^*(1430)\rightarrow \pi K$ &  $251.1$   \\
\hline
$~f_0(500)\rightarrow \pi \pi$ & Not conclusive  & $~f_0(1370)\rightarrow \pi \pi$ &  Not conclusive   \\
\hline
$~f_0(980)\rightarrow \pi \pi$ & $50$  & $~f_0(1500)\rightarrow \pi \pi$ &  $38.1$   \\
$~f_0(980)\rightarrow K\bar{K}$ & $9.5\text{--}46.2$  & $~f_0(1500)\rightarrow K\bar{K}$ & $9.5$  \\
\hline
\end{tabular}
\end{table}
Table~\ref{partial width} shows the experimental partial decay widths extracted from PDG 2022~\cite{PDG22}.
Except for $a_0(980)$, $a_0(1450)$ decaying into $K\bar{K}$, there seems to be a general tendency that
the partial decay width of the light nonet is larger than that of the heavy nonet,
\begin{eqnarray}
&&\Gamma_{exp}(\text{light nonet}) \geq \Gamma_{exp}(\text{heavy nonet})\label{trend}\ ,
\end{eqnarray}
For $f_0(500)$, $f_0(1370)$, their partial widths for the modes, $f_0(500)\rightarrow \pi \pi$, $f_0(1370)\rightarrow \pi \pi$, are not conclusive
from the present PDG data.
But the tendency like Eq.~(\ref{trend}) is still expected to hold in these cases as well.
The $f_0(500)$ is a resonance famous for its broad width and the $\pi\pi$ mode probably represents the full width of $f_0(500)$.
But for the $f_0(1370)$, the $\pi\pi$ mode is just one mode among various modes and the total width of $f_0(1370)$ is shared by all the decay modes.
Even though we do not have specific numbers for their widths, it is quite likely that their partial widths
also follow the general trend of Eq.~(\ref{trend}).

The general tendency represented by Eq.~(\ref{trend}) is not natural because,
kinematically, heavy resonances are expected to have larger partial widths as in Eq.~(\ref{kin}).
What is interesting is that this unnatural trend of Eq.~(\ref{trend}) can be explained if we have
$|G|\gg|G^\prime|$ as predicted by the tetraquark mixing framework.
Since partial decay widths are given by Eq.~(\ref{partial}), the tendency of experimental partial widths, Eq.~(\ref{trend}),
between the two nonets must be reproduced if we multiply $(\text{coupling strength})^2$ on both sides of Eq.~(\ref{kin}).
That is, multiplying $G^2$ on the left-hand side and $G^{\prime 2}$ on the right-hand side of Eq.~(\ref{kin}), we should have
\begin{eqnarray}
G^2 \Gamma_{kin} \text{(light~nonet)} \geq G^{\prime 2} \Gamma_{kin} \text{(heavy~nonet)} \label{upshot}\ ,
\end{eqnarray}
with the opposite inequality from Eq.~(\ref{kin}) in order to reproduce the tendency in the experimental
partial width, Eq.~(\ref{trend}).
Only way to get this opposite inequality from Eq.~(\ref{kin}) is to have $|G|\gg|G^\prime|$ as predicted by the tetraquark mixing framework.

One exceptional case is the isovector resonances decaying into $K\bar{K}$. In this case,
$\Gamma_{exp}[a_0(980)\rightarrow K\bar{K}] < \Gamma_{exp}[a_0(1450)\rightarrow K\bar{K}]$ so this does not follow the general trend
of Eq.~(\ref{trend}).
But, $\Gamma_{exp}[a_0(980)\rightarrow K\bar{K}]$ is smaller
by the kinematical cutoff. The central mass of $a_0(980)$, $\sim 980$ MeV, is smaller than the $K\bar{K}$ threshold, $\sim 990$ MeV
so the $a_0(980)$ can decay into $K\bar{K}$ only through high tail of the mass distribution broaden by the total width.
Even if this decay mode is amplified by the coupling strength, more than half of the partial width is blocked by the kinematical cutoff.
This is in contrast to the heavy nonet case, $a_0(1450)\rightarrow K\bar{K}$, where all the mass region of $a_0(1450)$ broadened by the total width
can decay to $K\bar{K}$ without suffering from the kinematical cutoff.
In fact, it can be demonstrated quantitatively that the partial widths explicitly calculated including the
mass distribution
as well as the coupling strengths given by Table~\ref{coup}
agree reasonably well with the experimental partial widths~\cite{Kim:2017yur,Kim:2022qfj}.
So this exceptional case does not undermine our conclusion, Eq.~(\ref{gg}).

In conclusion, our prediction from the tetraquark mixing model, $|G|(\text{light nonet})\gg|G^\prime|(\text{heavy nonet})$, is clearly supported by
the experimental partial decay widths satisfying $\Gamma_{exp}(\text{light nonet}) \geq \Gamma_{exp}(\text{heavy nonet})$.
Kinematically, the inequality in the experimental widths is very unnatural and
the tetraquark mixing model is likely the only one that can explain this unnatural trend.
The enhanced coupling strength in the light nonet also helps in part to understand why the $K^*_0(700)$ and $f_0(500)$ have very broad decay widths.
Since our prediction crucially relies on the fact that two types of tetraquark mix each other, the two nonets in PDG cannot be treated separately
when their physical properties are investigated.
Our predictions on the couplings hold for all members of the two nonets.
This contrasts with non-tetraquark models, where applications are often limited to some members of the two nonets.
Non-tetraquark models include meson molecular picture~\cite{Weinstein:1990gu}, $q\bar{q}$ picture with hadronic intermediate
states~\cite{Tornqvist:1995kr,Boglione:2002vv} and so on.
In this sense, it is quite unlikely that this prediction on the couplings can be reproduced from those models other than tetraquarks.
Therefore, our results strongly support that the two nonets in $J^{PC}=0^{++}$,
the light nonet [$a_0(980)$, $K_0^*(700)$, $f_0(500)$, $f_0(980)$] and the heavy nonet
[$a_0(1450)$, $K_0^*(1430)$, $f_0(1370)$, $f_0(1500)$], are tetraquarks generated from the admixture of two tetraquark types, $|000\rangle$ and $|011\rangle$.

\section*{Acknowledgments}
This work was supported by the National Research Foundation of Korea(NRF) grant funded by the
Korea government(MSIT) (No. NRF-2023R1A2C1002541, No. NRF-2018R1A5A1025563).

\end{document}